\documentclass[a4paper,10pt]{revtex4}
\usepackage{amssymb}
\usepackage{color}
\usepackage{graphicx}% Include figure files
\usepackage{subfig}
\usepackage{appendix}
\usepackage{bm}% bold math
\usepackage{CJKutf8}
\newcommand{\beq}{\begin{eqnarray}}
\newcommand{\eeq}{\end{eqnarray}}
\providecommand{\ignore}[1]{}

\begin{document}
\begin{CJK*}{UTF8}{bsmi}

\title{Coherently amplifying photon production from vacuum with a dense cloud of accelerating photodetectors}

%\author{Hui Wang}\affiliation{Department of Physics and Astronomy, Dartmouth College, Hanover, New Hampshire
%03755, USA }
%\email[]{hui.wang-2.gr@dartmouth.edu}

%\author{Miles Blencowe}\affiliation{Department of Physics and Astronomy, Dartmouth College, Hanover, New Hampshire
%03755, USA }

\author{Hui Wang}
\email[]{hui.wang-2.gr@dartmouth.edu}

\author{Miles Blencowe}

\address{Department of Physics and Astronomy, Dartmouth College, Hanover, New Hampshire
03755, USA }
\begin{abstract}
\noindent{\large\textbf{Abstract}}\\
An accelerating photodetector is predicted to see photons in the electromagnetic vacuum. However, the extreme accelerations required have prevented the direct experimental verification of this quantum vacuum effect. In this work, we consider many accelerating photodetectors  that are contained within an electromagnetic cavity.  We show that the resulting photon production from the cavity vacuum can be collectively enhanced such as to be measurable. The combined cavity-photodetectors system maps onto a parametrically driven Dicke-type model; when the detector number exceeds a certain critical value, the vacuum photon production undergoes a phase transition from a normal phase to an enhanced superradiant-like, inverted lasing phase. Such a model may be realized as a mechanical membrane with a dense concentration of optically active defects undergoing gigahertz flexural motion within a superconducting microwave cavity. We provide estimates suggesting that recent related experimental devices are close to demonstrating this inverted, vacuum photon lasing phase.
\end{abstract}

\maketitle
\end{CJK*}
\noindent{\large\textbf{Introduction}}\\
One of the most striking consequences of the interplay between relativity and the uncertainty principle is the predicted detection of real photons from the quantum electromagnetic field vacuum by non-inertial, accelerating photodetectors. For the special case of a photodetector with uniform, linear acceleration $a$ in Minkowski vacuum, the Unruh effect predicts that the detected photons are furthermore in a thermal state with temperature $T=\hbar a /(2\pi c k_{\mathrm{B}})$~\cite{Unruh, Dewitt,Crispino,Fulling,Ben}. However, to measure for example a $1~{\mathrm{K}}$ thermal photon temperature, a uniform detector acceleration $a=2.47\times10^{20}~ {\mathrm{m}}\, {\mathrm{s}}^{-2}$ is required, which seems impossibly high for any current or planned tabletop experiments. 

Given the difficulty in achieving  uniform photodetector accelerations of sufficient magnitude and duration, certain experimental approaches have instead considered electrons that are accelerated nonuniformly by intense electromagnetic fields such that the trajectories are confined to a finite spatial volume. In particular, Bell and Leinass~\cite{Bell} interpreted the observed spin depolarization of electron(s) undergoing uniform circular motion in a storage ring in terms of  effective heating by vacuum fluctuations  (see also Refs. \cite{Unruh98,Biermann}). And Schutzhold \emph{et al.}~\cite{Schutzhold} showed that oscillating electrons emit both classical (Larmor) and quantum radiation, with the latter corresponding to the accelerating electrons converting quantum vacuum fluctuations into entangled photon pairs. 

Motivated by the goal to enhance the detection of photons from vacuum, a proposal by Scully \emph{et al.}~\cite{Scully} locates an accelerating photodetector atom within an electromagnetic cavity; the idealized, perfectly conducting walls of the enclosing cavity effectively modify the spacetime geometry from Minkowki space to a finite volume for the electromagnetic vacuum~\cite{Dewitt}. This finite confining cavity volume then necessitates a non-uniform, i.e., oscillatory acceleration, similar to the accelerating electron proposal of Ref.~\cite{Schutzhold}; the photodetector atom is envisaged for example as attached to a piezoelectrically driven, vibrating cantilever structure. Another recent relevant proposal \cite{Lochan} considers a rapidly rotating, initially excited atom in a cavity, such that the atom's spontaneous emission is  enhanced when the sum of the rotation and atom splitting frequencies resonantly matches a  cavity mode frequency.   

In these investigations, the electrons and atom photodetectors accelerate nonuniformly (in either direction or magnitude) and in contrast to the Unruh effect as conventionally defined~\cite{Crispino,Fulling,Ben},  the predicted photon spectrum is not expressible as a thermal distribution with temperature given solely in terms of the acceleration and fundamental constants \cite{Hu,Doukas}. Nevertheless, the effects of entangled photon pair production and photon detection from vacuum are still direct consequences of the non-inertial motion of electrons and atoms.

In recent work~\cite{Wang,Miles}, we proposed close analogs of the vacuum, oscillatory photodetection effect that involve coplanar, i.e., two-dimensional (2D) superconducting microwave cavity circuit systems~\cite{nation2012}. The photodetector is modelled alternatively as a  harmonic oscillator~\cite{Wang} and as a qubit~\cite{Miles} that capacitively couple to the 2D cavity via a mechanically oscillating film bulk acoustic resonator (FBAR)~\cite{sanz2017}; by increasing the FBAR capacitor plate area, a strong detector-cavity coupling can be achieved. Furthermore, when the FBAR mechanical frequency matches the sum of the cavity mode and detector ground-first excited energy level splitting, the photon production rate from cavity vacuum is resonantly enhanced~\cite{Scully} so as to be detectable. The use of an FBAR introduces an actual mechanical acceleration, in contrast to previous circuit-based analogs where the accelerating photodetector is mimicked by an electromagnetically induced, time changing detector coupling~\cite{Felicetti, garcia2017}.

However, a feasible way to demonstrate ``genuine" oscillatory acceleration detection and production of photons from vacuum, where the photodetectors are accelerating within a three-dimensional (3D) electromagnetic cavity (as opposed to the above 2D analogues) is still lacking. Despite the resonant gain resulting from placing an accelerating atom inside a  cavity~\cite{Scully,Lochan}, currently achievable cavity quality factors, atom-cavity coupling strengths, and mechanical oscillation acceleration magnitudes are insufficient for attaining measurable photon production from vacuum for a single accelerating detector.

In this paper, we describe a way to realize acceleration photodetection from vacuum with a scheme that involves ``many" oscillating photodetectors contained within a microwave cavity (Fig. \ref{syst}). Such a scheme may be realized by exploiting optically active defects that are embedded within a vibrating membrane structure, such that the defects effectively function as accelerating photodetectors. As one concrete example, we consider a gigahertz (GHz) vibrating diamond membrane containing nitrogen vacancy (NV) color centre defects~\cite{Piracha}. We shall take advantage of a Dicke superradiance-like~\cite{Dicke,Gross,angerer2016,angerer2018}, vacuum amplification effect due to the collective motion of the many accelerating photodetectors. While the possibility has been mentioned before to coherently enhance photon production from vacuum using many accelerating electrons~\cite{Schutzhold} or detectors~\cite{delrey12, delrey15},  no such  investigations have  been carried out previously to the best of our knowledge. (Note, however, the  proposal of Ref. \cite{Kim} to amplify and hence detect cavity photons produced via the dynamical Casimir effect for an accelerating cavity mirror through their superradiant-enhanced interaction with a cloud of optically pumped atoms.) 
    
\vspace{0.5cm}
%We then present numerical and approximate analytical solutions to the model for various sample parameters, establishing the accuracy of the Dicke Hamiltonian description, as well as the existence of a phase transition in the vacuum photon production to a superradiant-like (properly termed `inverted lasing') phase. Finally, we discuss the possible realization involving a GHz oscillating diamond membrane containing NV center defects~\cite{Piracha,angerer2016,angerer2018} functioning as TLS photodetectors enclosed within a 3D microwave cavity.
\noindent{\large\textbf{Results}}\\
\noindent{\textbf{Cavity-$N$ accelerating detectors model}}
We first introduce our model for a cavity field interacting with $N\gg 1$ accelerating two level system (TLS) photodetectors and show how the model maps by approximation onto a simpler, parametrically driven Dicke-type Hamiltonian~\cite{Brandes2,chitra2015}. 
Referring to Fig. \ref{syst}, the $N$ TLS detectors are assumed to be embedded within a two-dimensional membrane material system undergoing driven, small amplitude GHz flexural vibrations in the $x$-coordinate direction (i.e., displacements normal to the static, equilibrium membrane $y$-$z$ surface), hence imparting acceleratory motion to the TLSs' centres of mass; the GHz membrane flexural motion may be actuated via piezoelectric transducers, for example. The membrane is enclosed by a microwave cavity `box' with volume $V=L_x L_y L_z$ and box dimensions $L_i\sim$ few cm.   The cavity dimension in the flexural motion ($x$ coordinate) direction is denoted as $L_x$, and the membrane is assumed to be located at the cavity midway point $x=L_x/2$. We consider the following starting model action for the combined cavity field-$N$ detector system:
\begin{eqnarray} 
S & = & -\int_V d^{3+1}x \frac{1}{2} \partial_{\mu} \Phi\partial^{\mu}\Phi + \sum_{i=1}^N \int d\tau_i \left\{\frac{m_0}{2} \left[\left(\frac{d Q_i}{d\tau_i}\right)^2 - \omega_{{\mathrm{d}}0}^2Q_i^2 \right] -\frac{g}{4!} Q_i^4 \right\}\cr
&&+ \lambda_0 \sum_ {i=1}^N \int d\tau_i \int_V d^{3+1}x Q_i(\tau_i) \Phi(t,\vec{r}) \delta^{3+1} \left[x^{\mu}-z_i^{\mu}(\tau_i)\right],
\label{action}
\end{eqnarray}
where the detectors, classically modelled by internal, one dimensional anharmonic oscillator coordinates $Q_i, \, i=1,\dots N$, are linearly coupled to a $3+1$ dimensional massless scalar field denoted as $\Phi(t,\vec{r})$ that models the electromagnetic field within the cavity.
Equation (\ref{action}) generalizes the starting action of Ref. \cite{Lin} to include more than one detector, as well as assumed weakly anharmonic oscillator potential energy terms $gQ_i^4/4!$.
The latter allows the detectors' internal degrees of freedom to be truncated to TLSs under parametric resonance conditions; while the harmonic oscillator detector model is beneficial for analytical calculations~\cite{Lin}, it is somewhat artificial since no detector is purely harmonic, and in particular results in an unphysical parametric instability beyond a critical field-detector(s) coupling~\cite{Wang}.
The detectors are assumed to have identical centre of mass rest frame internal harmonic oscillator frequency $\omega_{{\mathrm{d}}0}$, anharmonic coupling $g$, and coupling strength $\lambda_0$ between each detector's internal degree of freedom and the cavity field. Possible, direct interaction terms between the detectors are not considered in our model; as we show later below, coherent enhancements in the photon production from vacuum can occur as a consequence of the detectors coupling via the cavity field, provided the average spacing between neighbouring detectors is much smaller than the resonant cavity field mode wavelength.

With the membrane undergoing driven flexural oscillations at some frequency $\Omega_{\mathrm{m}}$, the centre of mass of the $i$th detector follows the worldline $z^\mu_i(t) = (t, L_x/2 +A \cos\left(\Omega_{\mathrm{m}} t+\phi_i\right), y_i, z_i)$, where $A$ is the detector's centre of mass acceleration amplitude and $\phi_i$ its  phase. The phase $\phi_i$ accounts for the fact that the membrane's GHz flexural mode wavelength ($\sim$ few $\mu$m) is much smaller than the extent of the detector distribution ($\gtrsim$ few mm), so that individual oscillating detectors may be out of phase with respect to each other depending on their relative separation in the $y$-$z$ plane. The $i$th detector's proper time is denoted by $\tau_i$, and its transverse centre of mass coordinates are assumed to satisfy $y_i\sim L_y/2$ and $z_i\sim L_z/2$. The latter condition allows the simplifying approximation that the transversely distributed detectors couple equally to the resonant cavity mode. Note that we adopt the Minkowski metric sign convention $\eta_{\mu\nu}=\mathrm{diag}(-1,1,1,1)$.

The action (\ref{action}) yields the following model Hamiltonian for the cavity-TLS detectors system in the rest frame of the cavity as derived in Supplementary Note 1:
\begin{eqnarray} 
H & = & \hbar \omega_{\mathrm{c}} a^{\dag} a + \hbar \tilde{\omega}_{\mathrm{d}} \sum_ {i=1}^N \frac{d\tau_i} {dt} \frac{\sigma_i^z}{2} + \hbar \tilde{\lambda} \sum_ {i=1}^N \frac{d\tau_i} {dt} \sin \left[k_\mathrm{c} A \cos(\Omega_{\mathrm{m}} t + \phi_i) \right] ({a} + {a}^{\dag}) \sigma_i^x,
\label{hamilton}
\end{eqnarray}
where we assume a single mode approximation for the cavity field   with mode frequency $\omega_{\mathrm{c}}=c k_\mathrm{c}$, $k_\mathrm{c}=\sqrt{(2\pi/L_x)^2+(\pi/L_y)^2+(\pi/L_z)^2}$, and we truncate the anharmonic oscillator detector state space to that spanned by the ground and first excited energy eigenstates with transition frequency $\tilde{\omega}_{\mathrm{d}}$  and with TLS-cavity mode coupling $\tilde{\lambda}$ defined in Supplementary Note 1. Here, we consider the $(n_x,n_y,n_z)=(2,1,1)$ cavity mode, which has a node at $L_x/2$ (see Fig. \ref{syst}) resulting in a first order dependence on oscillation amplitude $A$ in the interaction term of Hamiltonian (\ref{hamilton}); if we were to instead use the $(n_x,n_y,n_z)=(1,1,1)$ mode, the coupling would  involve a cosine instead of a sine term and hence would be much smaller at second order in $A$. The cavity single mode approximation and detector two level truncation are justified under the condition of parametric resonance between the driven, flexural membrane frequency $\Omega_{\mathrm{m}}$ and the (renormalized) detector and cavity mode frequencies (see later below). Hamiltonian (\ref{hamilton}) accounts for relativistic time dilation for the accelerating TLSs through the presence of the reciprocal Lorentz factors ${d\tau_i}/ {dt} = \sqrt{1-\xi^2 \sin^2 \left(\Omega_{\mathrm{m}} t+\phi_i\right)}$, where $\xi = \Omega_{\mathrm{m}} A /c$,  and may be thought of as a relativistic, parametrically driven Dicke model \cite{Brandes2,chitra2015}. 

With for example an achievable  membrane flexural, microwave scale frequency $\Omega_{\mathrm{m}} \sim 2\pi \times 10~{\mathrm{GHz}}$ and oscillation amplitude $A\sim 10^{-10}~{\mathrm{m}}$, we have $\xi= \Omega_{\mathrm{m}} A /c\sim 10^{-8}$ and hence the time dilation factors can be safely neglected for potential laboratory realizations. The sinusoidal interaction term can then be well-approximated as $\sin \left[k_\mathrm{c} A \cos\left(\Omega_{\mathrm{m}} t+\phi_i\right)\right]\approx k_\mathrm{c} A \cos\left(\Omega_{\mathrm{m}} t+\phi_i\right)=\omega_{\mathrm{c}} A/c \cos\left(\Omega_{\mathrm{m}} t+\phi_i\right)$, and with $\omega_{\mathrm{c}}\sim\Omega_{\mathrm{m}}$ under conditions of resonance, we see that the individual TLS detector-cavity coupling $\tilde{\lambda}$ is effectively reduced by the maximum TLS velocity-to-speed of light ratio $v_{\mathrm{m}}/c\ll 1$. 

Nevertheless, from a theoretical standpoint it is still interesting to allow also for relativistic TLS accelerations in exploring photon production when starting initially with the TLSs in their ground states and the cavity in its vacuum state. 
As we show just below, even under conditions of extreme relativistic TLS motion, Hamiltonian $(\ref{hamilton})$ can be mapped onto a much simpler, time-independent Hamiltonian  through a type of renormalized, rotating wave approximation (RWA), provided the TLS-cavity coupling $\tilde{\lambda}$ is sufficiently small (see Supplementary Note 1 for the details of the mapping).   

In particular, setting the phases $\phi_i=0$, we can describe the collection of $N$ TLSs as a single $N +1$-level system viewed as a large pseudospin vector of length $j=N/2$, with the collective spin operators given by $J^z \equiv \sum_{i=1}^N {\sigma_i^z}/{2}$ and $J^{\pm} \equiv \sum_{i=1}^N \sigma_i^{\pm}$. Expanding the time-dependent terms in Eq.~(\ref{hamilton}) and keeping only terms up to second harmonics in $\Omega_{\mathrm{m}}$, the TLS transition frequency is renormalized as  $\omega_{\mathrm{d}}=\tilde{\omega}_{\mathrm{d}} D_0$ and there is an additive frequency modulation ${\tilde{\omega}_{\mathrm{d}}D_2} \cos(2\Omega_{\mathrm{m}}t)$, where $D_0$ and $D_2$ are $\xi$-dependent constants (defined in Supplementary Note 1). Imposing the resonance condition $\Omega_{\mathrm{m}} = \omega_{\mathrm{c}} + \omega_{\mathrm{d}}$, we then transform the frequency renormalized, modulated Hamiltonian  to the rotating frame via the unitary operator $U_{\mathrm{RF}}(t) = \exp\left(i\omega_{\mathrm{c}}a^{\dag}at + i J^z \left[\omega_{\mathrm{d}}t + {\tilde{\omega}_\mathrm{d}D_2} \sin(2\Omega_{\mathrm{m}}t)/ {2\Omega_{\mathrm{m}}} \right]\right)$. Applying the rotating wave approximation (RWA), we finally have that the relativistic, driven Dicke type Hamiltonian (\ref{hamilton}) can be replaced by  the following much simpler approximate Hamiltonian:
\begin{eqnarray} 
H & \approx & \hbar\lambda (a^{\dag} J^+ + a J^-),
\label{spinrwa}
\end{eqnarray}
where the coupling constant is $\lambda=\frac{1}{2} \tilde{\lambda} C_1 \left[J_0 \left ({\tilde{\omega}_\mathrm{d}D_2} /{2\Omega_{\mathrm{m}}}\right) -J_1 \left({\tilde{\omega}_{\mathrm{d}} D_2} /{2\Omega_{\mathrm{m}}}\right)\right]$, with $C_1$ a $\xi,\, \Omega_{\mathrm{m}}$-dependent constant (defined in Supplementary Note 1), and $J_0(z)$, $J_1(z)$ are Bessel functions of the first kind (not to be confused with the spin operators $J^{\pm}$).

An actual microwave cavity will be lossy, while TLSs in their excited state will relax through photon and phonon emission, with the latter decay channel a consequence of the TLSs embedded within an elastic membrane. For the possible realization considered below, the dominant TLS relaxation channel is through phonon emission~\cite{angerer2016,angerer2018}, and since as mentioned above the phonon wavelength is much smaller than the extent of the TLS distribution, we model the TLSs as coupled to approximately independent environments. We assume that the cavity-TLSs system dynamics can be described by the following Lindblad master equation: \begin{eqnarray}
\dot{\rho} = -\frac{i}{\hbar} [H,\rho] + \gamma_{\mathrm{c}} \mathcal{L}_{a}[\rho] + \gamma_{\mathrm{d}} \sum_{i=1}^N \mathcal{L}_{\sigma_i^-} [\rho],
\label{lindblad}
\end{eqnarray}
where $\gamma_{\mathrm{c}}$ and $\gamma_{\mathrm{d}}$ are the cavity and individual TLS energy damping rates,  respectively,
and the Lindblad superoperator is defined as $\mathcal{L}_{A} [\rho] \equiv A\rho A^{\dag} - \frac{1}{2} A^{\dag}A\rho - \frac{1}{2} \rho A^{\dag}A$. Here,  we suppose that the environment temperature is small compared to the frequencies of the cavity mode and TLSs (i.e., $k_\mathrm{B} T/\hbar\ll \omega_{\mathrm{c}},\,\omega_{\mathrm{d}}$), and also that each TLS has approximately the same damping rate; for the possible realization with microwave cavity and detector frequencies of a few GHz, it suffices to work at dilution fridge temperatures of a few tens of mK or below in order to be in this low temperature regime.  

\noindent{\textbf{Analysis of the model}}
The single cavity mode approximation in Hamiltonian (\ref{hamilton}) is justified provided the approximate resonance condition $\Omega_{\mathrm{m}} \approx \omega_{\mathrm{c}} + \omega_{\mathrm{d}}$ is satisfied~\cite{Wang}. This resonance condition also justifies the RWA Hamiltonian (\ref{spinrwa}), provided $|\lambda|\ll\omega_{\mathrm{c}},\,\omega_{\mathrm{d}}$. The advantage to working with the much simpler Hamiltonian (\ref{spinrwa}) is that the Lindblad master equation (\ref{lindblad}) can be solved numerically for up to 20 or so TLSs, while for the full, relativistic time-dependent Hamiltonian (\ref{hamilton}), numerically solving the Lindblad equation is only possible for less than around 10 TLSs before the computational run times become excessively long.   It is noteworthy in this respect that    using the approximate Hamiltonian $(\ref{spinrwa})$ in the master equation provides an accurate description of the average photon number dynamics even for relativistic TLS centre of mass motions with $\xi= \Omega_{\mathrm{m}} A /c\lesssim 1$  (see Supplementary Figure 1). Furthermore, for the assumed independent TLS damping, the average photon number dynamics is relatively insensitive to the phase $\phi_i$-dependencies of the TLSs' motions, justifying setting $\phi_i=0$ above (see Supplementary Figure 1). From now on, we will use Hamiltonian (\ref{spinrwa}) in the master equation (\ref{lindblad}).  

One analyical approach to solving the above master equation (but with independent damping replaced by collective damping \cite{chitra2015}--see below) begins with the Holstein-Primakoff (H-P) transformation $J^+ = b^{\dag} \sqrt{2j - b^{\dag}b}$, $J^- = \sqrt{2j - b^{\dag}b} b$, which maps the collective spin operators $J^\pm$ to the bosonic creation and annihilation operators $b^{\dag}$, $b$  \cite{Holstein}. In the large $j=N/2$ limit, the collective spin of the $N$ TLSs is approximated as a single harmonic oscillator, with Hamiltonian  (\ref{spinrwa}) mapped to $\hbar\sqrt{N} \lambda (a^{\dag} b^{\dag} + ab)$. This coincides with the RWA Hamiltonian that was used in Ref. \cite{Wang} to describe the oscillatory acceleration for a system comprising a photodetector coupled to a cavity electromagnetic field in the single mode approximation, with the detector's internal degrees of freedom modeled as a harmonic oscillator, and the detector's centre of mass oscillating at a frequency $\Omega_{\mathrm{m}}$ that matches the sum of the cavity frequency and detector internal frequency. This system describes a nondegenerate parametric amplification process, with cavity-detector photon pairs produced from the vacuum via mechanical pumping. Including cavity and detector damping and solving analytically the corresponding master equation for the above bosonic Hamiltonian, the average cavity photon number in the long time limit is $\langle a^{\dag}a\rangle = 4\gamma_{\mathrm{d}} N \lambda^2/[(\gamma_{\mathrm{c}} + \gamma_{\mathrm{d}})(\gamma_{\mathrm{c}} \gamma_{\mathrm{d}} -4N\lambda^2)]$. Note that the system exhibits a parametric instability when the effective coupling strength $\sqrt{N} \lambda$ exceeds the value $\sqrt {\gamma_{\mathrm{c}} \gamma_{\mathrm{d}}}/2$. While such an instability indicates a breakdown of the above H-P derived approximation method, it does point to the existence of an enhanced vacuum photon production phase for $N>N_{\mathrm{crit}}=\gamma_{\mathrm{c}}\gamma_{\mathrm{d}}/(4\lambda^2)$  in the original cavity-TLSs model dynamics given by equations (\ref{hamilton}--\ref{lindblad}), where there is no such instability \cite{Kirton3}. In the following analysis, we will continue to use the definition $N_{\mathrm{crit}}=\gamma_{\mathrm{c}}\gamma_{\mathrm{d}}/(4\lambda^2)$ also for the independent damping master equation (\ref{lindblad}); as we show in Supplementary Note 2, a cumulant expansion analysis of this equation  establishes that $N_{\mathrm{crit}}$ delineates different phases of photon production.

In Fig. 2a, we show for some example, illustrative parameters corresponding to $N_{\mathrm{crit}}=1$, the dynamical behavior of the average scaled cavity photon number  $\langle a^\dag a \rangle/N$ starting from the cavity mode vacuum and with all the TLSs initially in their ground states at time $t=0$. The solid line plots are obtained by numerically solving \cite{johansson2013} the Lindblad master equation (\ref{lindblad}) with the RWA Hamiltonian (\ref{spinrwa}); the utilized numerical method  exploits the permutation symmetry of the density operator~\cite{Shammah}, which reduces the size of the Hilbert space required for the TLSs and increases the accessible value of $N$ up to around 20, depending on the parameter choices. The dashed line plots are obtained by solving approximate equations for the non-vanishing first and second order moments derived from the master equation (\ref{lindblad}). We find dramatically improved accuracy as compared with usual cumulant expansion approximation methods \cite{Kirton1,Kirton2,Kirton3}, even for relatively small $N$,  by instead setting certain fourth order instead of third order cumulants to zero (see Supplementary Figure 1). In particular, with each of the  TLSs giving an identical contribution to the moment equations, we can replace $\sigma_i^z$, $\sigma_i^{\pm}\sigma_j^z$  ($i\neq j$) with $\sigma_1^z$, $\sigma_2^{\pm}\sigma_1^z$ respectively. Furthermore, utilizing the identity $\sigma_1^z = 2\sigma_1^+ \sigma_1^- -1$ and approximating that the fourth cumulants vanish, we obtain for example the following non-vanishing third moment approximation (see Supplementary Equation S21):
$\langle a^{\dag} \sigma_2^+ \sigma_1^z\rangle = \langle a^{\dag} \sigma_2^+\rangle \langle \sigma_1^z \rangle + 2\langle a^{\dag} \sigma_1^+\rangle \langle \sigma_1^- \sigma_2^+\rangle$. Note that the latter approximation contains an additional term involving products of second moments as compared with the usually employed, third cumulant vanishing approximation~\cite{Kirton1,Kirton2,Kirton3}. 

The purpose for using the above cumulant expansion approximation is that the photon production from cavity vacuum can be investigated for $N\ggg 1$, relevant for possible realizations, while the numerical Lindblad solutions for smaller $N$ are useful for validating the cumulant approximation, as well as giving a more complete picture of the quantum dynamics.   

As $N$ increases above $N_{\mathrm{crit}}$, a burst peak of cavity photon production from vacuum appears in Fig. 2a that progressively grows in magnitude, narrows, and shifts to earlier times. Furthermore, the long time limit steady state average photon number  grows in magnitude. These features qualitatively resemble those of Dicke superradiance~\cite{Dicke,Gross}, although in the latter process the TLSs are initially prepared in their excited state and the cavity in the ground state. (Note that Ref. \cite{richter2017} considers the effect on superradiance of $N=6$ uniformly accelerating, initially excited photodetectors, which is  different from the collective enhancement effect for photon detection from vacuum considered here.) 

In order to get a better idea of the cavity mode-TLSs state, in Fig. 2b we show  for $N=15$ the time dependence of the logarithmic negativity measure of entanglement $E_{\mathcal{N}}$ {\cite{Vidal}}  between the cavity mode and TLS ensemble and also the cavity mode state Fano factor $(\langle(a^{\dag}a)^2\rangle-\langle a^{\dag}a\rangle^2)/\langle a^{\dag}a\rangle$ in comparison with the average scaled cavity photon number. The entanglement grows and reaches a maximum roughly when the Fano factor is a maximum, while the entanglement is relatively small, although non-zero, in the long-time limit. This non-zero entanglement growth from vacuum indicates that the cavity photon production and TLS excitation are correlated. Such entanglement probes are essential in potential future experiments in order to distinguish correlated vacuum photon production and TLS excitation from possible radiative heating effects (for which the entanglement between the cavity mode and TLSs vanishes).  The Fano factor gives partial information about the cavity mode state, in particular the photon number variance, and complements the information provided by the cavity mode Wigner function snapshots shown in Fig. 2c;  these latter  snapshots correspond to the instants when the Fano factor reaches its peak, subsequent trough, and long-time limit steady state. The photon production from a vacuum burst corresponds to the cavity mode  Wigner function rapidly first spreading out and then forming a ring with width close to that of a coherent state (characterized by Fano factor $=1$), and eventually settling into a thicker ring in the steady state due to environmental diffusion (with Fano factor $>1$). The Fano factor and Wigner function plots help to characterize, for example, how close the cavity mode state is to an effective thermal state (i.e., having a centred, Gaussian Wigner function with Fano factor $>1$); in contrast to the usual Unruh effect involving a uniformly accelerating detector, we see from Fig. 2c that the vacuum generated, cavity photon state with its ring-like Wigner function in the long-time limit is clearly non-thermal. This is not surprising given that the TLSs' motion is non-uniform, i.e., oscillatory.

\noindent{\textbf{Possible implementation}}
In order to get a more quantitative sense of the average cavity photon number dynamics scaling dependence on TLS number $N$, as well as model possible experimental set-ups, we must go to the large $N_{\mathrm{crit}}$ limit where it is not feasible to numerically solve the master equation (\ref{lindblad}). As an alternative, we can apply the approximate first and second order moment equations (see Supplementary Equation S22) that should become increasingly accurate  in the large $N$ limit, provided $N$ is not too close to $N_{\mathrm{crit}}$~\cite{Kirton1,Kirton2}, and which can be straightforwardly solved numerically for arbitrarily large $N$ since the number of coupled moment equations is fixed and small.

 We will assume in part the parameters of the 3D microwave cavity-coupled nitrogen vacancy (NV) centre defect scheme of Refs. \cite{angerer2016,angerer2018}, which observed signatures of Dicke superradiance \cite{Dicke,Gross}. In particular, we consider a coupling strength $\tilde{\lambda}=  2\pi\times 0.07~{\mathrm{Hz}}$, corresponding to an NV defect coupling via its magnetic moment to the magnetic field component of the considered cavity electromagnetic vacuum mode with frequency $\omega_{\mathrm{c}}=2\pi\times 3.2~{\mathrm{GHz}}$ \cite{angerer2018}. Assuming  a diamond membrane flexural oscillation amplitude $A=10^{-10}~{\mathrm{m}}$, the nonrelativistic coupling strength is $\lambda=\tilde{\lambda} \omega_{\mathrm{c}} A/(2c)\approx 1.5 \times 10^{-9}~{\mathrm{s}}^{-1}$. For the NV defects, the dominant relaxation process is through spin-phonon interactions with rate $\gamma_{\mathrm{d}}\approx 2\times 10^{-4}~ {\mathrm{s}}^{-1}$ \cite{angerer2018}. On the other hand, assuming a realizable, superconducting microwave cavity quality factor $Q_c=10^6$ \cite{stammeier2018}, we have $\gamma_{\mathrm{c}}=\omega_{\mathrm{c}}/Q_c\approx 2\times 10^4~{\mathrm{s}}^{-1}$. With these numbers, we have $N_{\mathrm{crit}} = {\gamma_{\mathrm{c}}\gamma_{\mathrm{d}}}/ \left({2\lambda}\right)^2 \approx 4\times 10^{17}$; the number $N$ of microwave field-coupled defects in the experiment of Ref. \cite{angerer2018} is in the range $(0.36-1.5)\times 10^{16}$, not far below this $N_{\mathrm{crit}}$ value.

Fig. 3a gives the first burst peak maximum and the long time limit steady state  rescaled cavity photon number $\langle a^\dag a \rangle/N$ dependencies on $N/N_{\mathrm{crit}}$.  We observe clear evidence of a phase transition, with the slope of average photon number dependence on $N$ changing sharply as $N$ moves through $N_{\mathrm{crit}}$. In particular, in the long time steady state, we obtain the following approximate analytical expressions well below and above $N_{\mathrm{crit}}$, respectively (see Supplementary Equation S24): $\langle a^{\dag} a\rangle=\frac{N}{N_{\mathrm{crit}}} \frac{\gamma_{\mathrm{d}}}{\gamma_{\mathrm{c}}+\gamma_{\mathrm{d}}}\approx 2 \times 10^{-26} N\lll 1$ for $N\ll N_{\mathrm{crit}}$, and $\langle a^{\dag} a\rangle=N{\gamma_{\mathrm{d}}}/{(2\gamma_{\mathrm{c}}})\approx 5 \times 10^{-9} N\ggg 1$ for $N\gg N_{\mathrm{crit}}$; we can clearly see that the average cavity photon number generated from vacuum is negligible below  and non-negligible above $N_{\mathrm{crit}}$. While these steady state average photon numbers scale linearly with $N$, from Fig.~\ref{lasing}b we see that in contrast the first burst peak scales as $N^\alpha$ with $\alpha \approx 2$ for $N>N_{\mathrm{crit}}$ (i.e., quadratically with $N$ to a good approximation); the peak average photon number rapidly grows in magnitude relative to the steady state value with increasing $N$ above $N_{\mathrm{crit}}$. In Fig. \ref{lasing}c, we see that the delay $t_{\mathrm{d}}$ in the appearance of this first burst peak starting from the initial cavity vacuum state becomes shorter as $N$ increases, with the inverse delay $t_{\mathrm{d}}^{-1}$ scaling linearly with $N$.

While these observed scaling dependencies for the first burst peak of photon production from vacuum coincide with those for Dicke superradiant bursts~\cite{angerer2018}, they are not properly a superradiant phase however \cite{Kirton2,Kirton3, Shchadilova}. In order to understand better the nature of this enhanced vacuum photon production phase, it is informative to apply the unitary transformation $U=\exp({i\pi J^x}/{2})$ to the master equation (\ref{lindblad}); the RWA Hamiltonian (\ref{spinrwa}) then transforms to the Tavis-Cummings Hamiltonian $H= \hbar\lambda (a^{\dag} J^- + a J^+)$, which involves co-rotating terms only, while the Lindblad operators $\mathcal{L}_{\sigma_i^-} [\rho]$ are transformed to $\mathcal{L}_{\sigma_i^+} [\rho]$, with each TLS initial ground state transformed to its corresponding excited state. The  cavity-oscillatory TLSs system is thus unitarily equivalent to a Tavis-Cummings model with an incoherent pump. While the latter model does not exhibit a Dicke superradiant transition, which requires the presence of both co-rotating and counter-rotating terms of the Hamiltonian, it nevertheless exhibits a so-called ``inverted lasing" (or ``counterlasing") transition \cite{Kirton2,Kirton3, Shchadilova}. For this reason, the transition to enhanced acceleration detection from vacuum may be thought of as a transition from a normal, incoherent phase below $N_{\mathrm{crit}}$ to a coherent inverted lasing phase above  $N_{\mathrm{crit}}$.

\vspace{0.5cm}
\noindent{\large\textbf{Discussion}}\\
The challenge of any experimental scheme is to exceed the threshold for coherent photon production from vacuum (equivalently the inverted lasing threshold \cite{Kirton2,Kirton3, Shchadilova}), which in terms of the TLS number $N$ is given by the condition $N >N_{\mathrm{crit}}= {\gamma_{\mathrm{c}}\gamma_{\mathrm{d}}}/ \left({2\lambda}\right)^2$. In order to reduce the size of $N_{\mathrm{crit}}$, we require microwave cavities with large quality factors, TLSs with small damping rates, and large TLS-photon coupling strengths ${\lambda}$. One way to enhance ${\lambda}$ is to reduce the microwave cavity volume, for example by utilizing 2D coplanar microwave cavities \cite{devoret2007,Probst,Ranjan}. However, this creates challenges for locating a sufficiently large number of TLSs within the microwave cavity region. On the other hand, while achievable ${\lambda}$ couplings for 3D microwave cavities are a few orders of magnitude smaller than what is possible for 2D coplanar cavities, correspondingly much greater numbers of TLSs can be located within the 3D microwave cavity region, furnished by defects within a flexurally vibrating membrane. The above mentioned NV centre scheme for probing Dicke superradiance~\cite{angerer2016,angerer2018} is a promising direction, although larger cavity quality factors are required, as well as larger surface area diamond membranes that can be actuated into GHz frequency flexural motion. 

While we have focused in the present investigation on a microwave cavity-GHz oscillating membrane defect scheme, it is worth considering other possible schemes as well that effectively incorporate many accelerating photodetector/emitter degrees of freedom. In particular, it would be interesting to revisit the oscillating electron-photon pair production scheme of Ref. \cite{Schutzhold}, but instead consider many electrons accelerating in a strong, periodic electromagnetic field. Under conditions where the dominant wavelength of emitted photon pairs is large compared to the average electron separation, we might expect a superradiant-like coherent enhancement of the pair production rate. Alternatively, we might consider a cloud of atoms (i.e., quantum dipoles)~\cite{zhu}, although the challenge to impart sufficient acceleration magnitude to the cloud's centre of mass would need to be addressed. Furthermore, electron-electron and dipole-dipole interactions would likely need to be taken into account; such interactions might result in quantum synchronization~\cite{zhu},  where small initial differences in the individual electron/dipole acceleration magnitudes and phases do not influence the resulting  coherent emission of radiation starting from the vacuum state. Finally, with the long-sought goal to demonstrate the original Unruh effect, it would be interesting to investigate possible collective enhancements of photodetection from vacuum for a dense cloud of detectors accelerating uniformly over a sufficiently long time interval in Minkowski vacuum, such that the detected photon spectrum is thermal to a good approximation.

In conclusion, we have proposed a way to demonstrate photon production from vacuum as a consequence of genuinely accelerating photodetectors. Our scheme involves a membrane with a dense cloud of embedded NV defects (equivalently TLS photodetectors) undergoing driven, transverse flexural vibrations within a microwave cavity, and modelled as a driven Dicke-type Hamiltonian. Under the condition of resonance, where the TLS's centre of mass acceleration frequency matches the sum of the cavity mode and internal TLS's transition frequency, the system Hamiltonian considerably simplifies via the RWA,  thus allowing for an accurate analytical solution using a cumulant expansion approach. When the number $N$ of TLS defects exceeds a critical value $N_\mathrm{crit}$, the system undergoes a transition to an inverted lasing phase, signalled by a `burst' peak in the average cavity photon number that scales as $N^2$, yielding significantly enhanced, collective photon production from vacuum. 

While the primary motivation for the present investigation concerns the interplay between the fundamental physics of relativistic quantum field vacuum and many body quantum dynamics, the possibility to realize entangled many TLS (i.e., qubit)-microwave photon states, purely by mechanically driving a membrane structure initially in a cavity vacuum, may also find broader relevance and application in quantum information science and technology.     

\vspace{0.5cm}
\noindent{\large\textbf{Data availability}}\\
The codes and generated data that were used to produce the figures in this paper are available by request  from the corresponding author.

\vspace{0.5cm}
\noindent{\large

\vspace{0.5cm}
\noindent{\large\textbf{Acknowledgements}}\\
We thank A. D. Armour, W. F. Braasch, S.-Y. Lin, O. B. Wright, N. Shammah, A. R. H. Smith, and J. Yang  for very helpful discussions. This work was supported by the NSF under grant nos. DMR-1507383 and PHY-2011382. 

\vspace{0.5cm}
\noindent{\large\textbf{Author contributions}}\\
H. W. and M. B. jointly conceived the idea for the project. H. W. performed the analytical and numerical calculations, and plotted the figures. H. W. and M. B. both contributed to the interpretation of the calculations and jointly wrote the paper.

\vspace{0.5cm}
\noindent{\large\textbf{Competing interests}}\\
The authors declare no competing interests.

\begin{figure}[ht]
\begin{center}
\includegraphics[width=3.5in,height=3.5in]{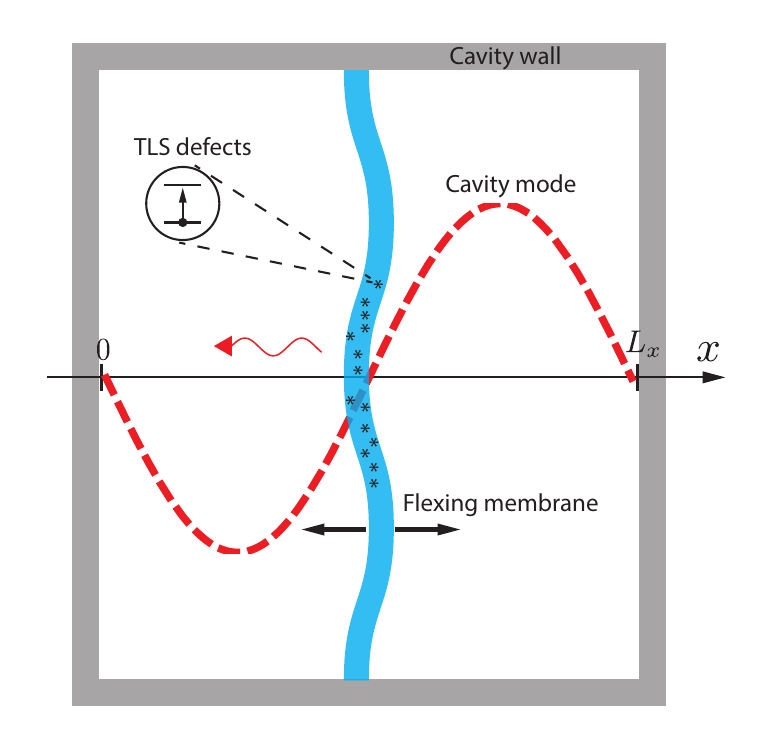} 
\caption{\label{syst}{\bf Cavity-accelerating photodetectors scheme}. A two-dimensional membrane (shown in crossection) located at $x=L_x/2$, midway between cavity wall (mirror) boundaries at $x=0, L_x$, undergoes flexural vibrations normal to its surface in the $x$-coordinate direction, resulting in the oscillatory acceleration of pointlike, two level system (TLS) photodetectors embedded within the membrane. An accelerating TLS can undergo a transition to its excited state level, with the accompanying emission of a photon into the cavity mode vacuum. A possible implementation might utilize a diamond membrane with nitrogen vacancy (NV) defect centres~\cite{Piracha,angerer2016,angerer2018} furnishing the TLSs, where the flexural vibrations of the membrane are induced for example by piezoactuators (not shown).}
\end{center}
\end{figure}

\begin{figure*}[ht]
\begin{center}
\includegraphics[width=0.98\textwidth]{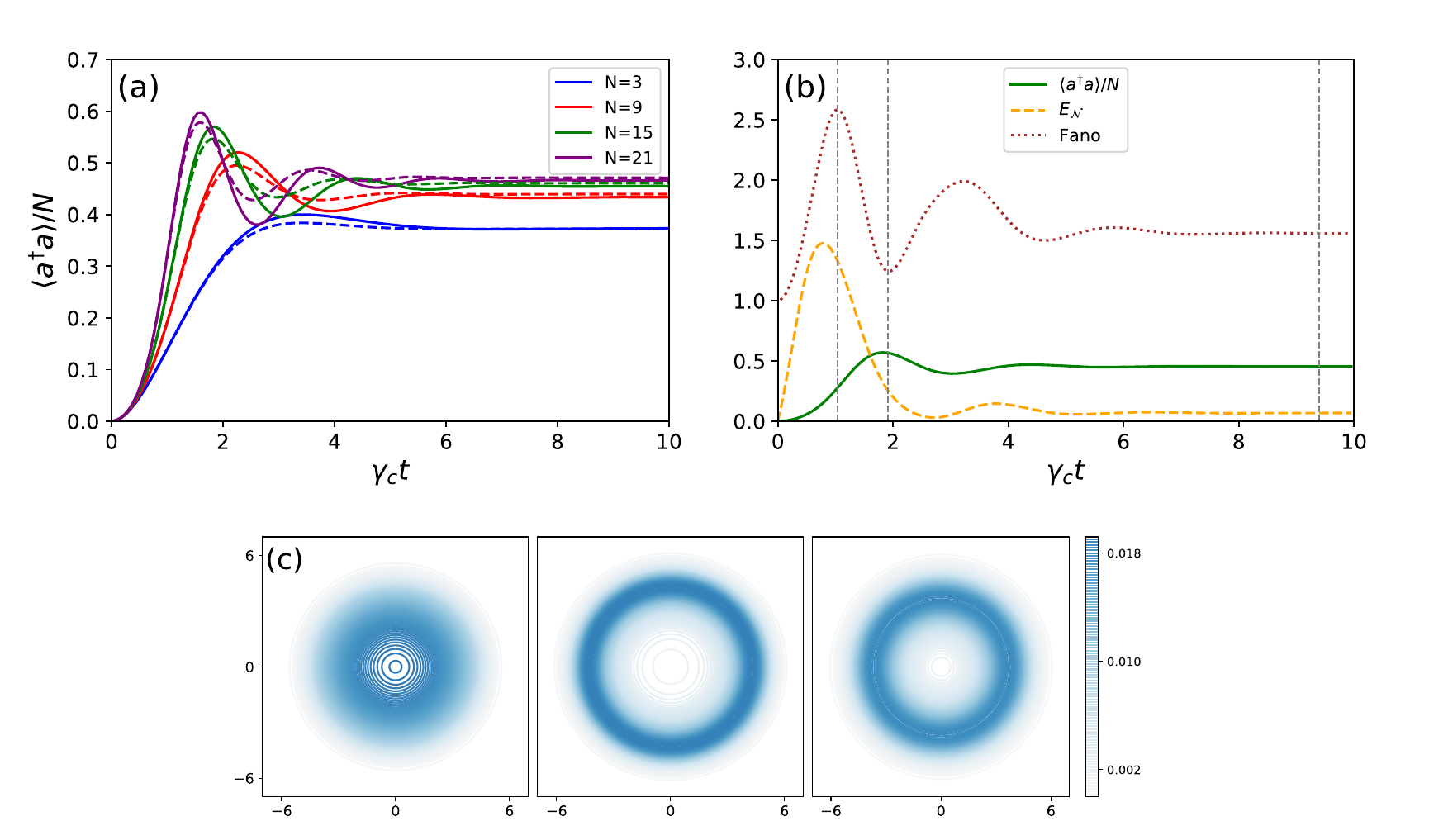} 
\caption{{\bf{Dynamics of the model.}} {\bf{a}} Evolution of the scaled cavity average photon number: $\langle a^\dag a \rangle/N$; solid lines are the numerical solutions and dashed lines are the cumulant expansion approximation. {\bf{b}} Evolution of $\langle a^\dag a \rangle/N$ (green solid line), Fano factor (purple dotted line), as well as the entanglement (logarithmic negativity $E_{\mathcal{N}}$ -- orange dashed line) for $N=15$ case. {\bf{c}} Wigner function snapshots at approximate time instants indicated by vertical dashed lines appearing in Fig. 2b. The parameters are $\gamma_{\mathrm{c}}=\gamma_{\mathrm{d}}=0.02$ and $\lambda=0.01$ (in units $\omega_{\mathrm{c}}=1$), corresponding to  $N_{\mathrm{crit}}=1$.}
\label{burst}
\end{center}
\end{figure*} 

\begin{figure}[ht]
\begin{center}
\includegraphics[width=0.98\textwidth]{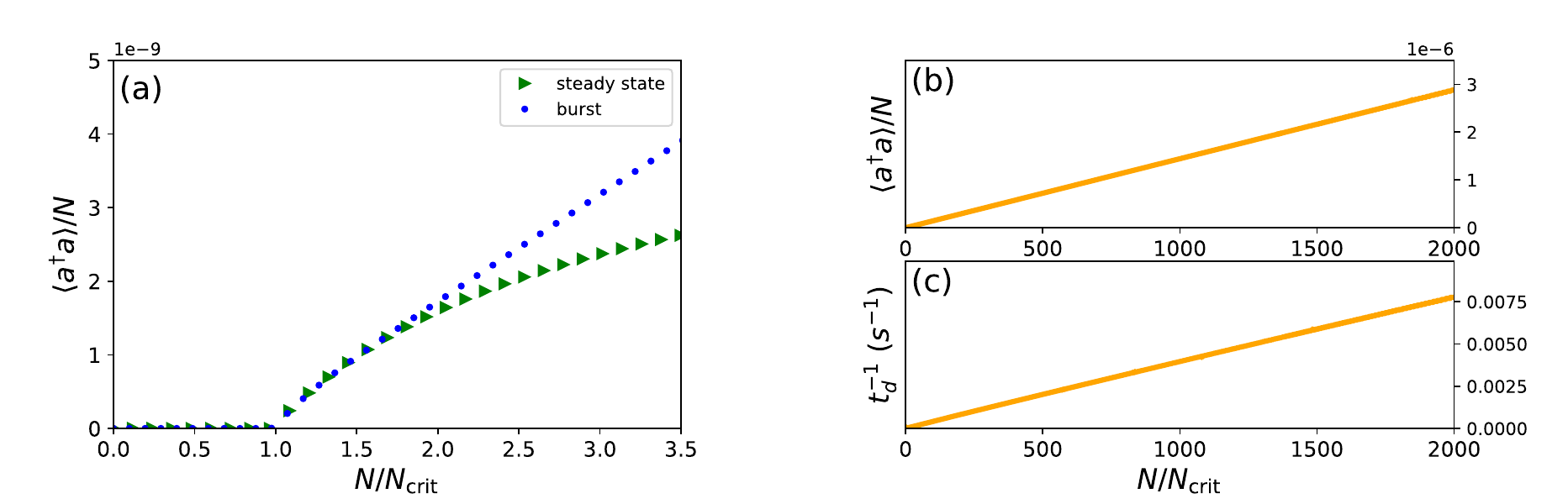} 
\caption{{\bf Photon production phase transition}. {\bf a} Scaled cavity photon number $\langle a^\dag a \rangle/N$ versus $N/N_\mathrm{crit}$ at the instant of the first vacuum photon production burst peak (blue dots)  and in the long time limit steady state (green triangles). {\bf b} First peak value of $\langle a^\dag a \rangle/N$ versus $N/N_\mathrm{crit}$  for $N\gg N_\mathrm{crit}$. {\bf{c}} First peak inverse delay time  $1/t_{\mathrm{d}}$ versus $N/N_\mathrm{crit}$  for $N\gg N_\mathrm{crit}$. Assumed parameters are $\gamma_{\mathrm{c}} \approx 2\times 10^4~{\mathrm{s}}^{-1}$, $\gamma_{\mathrm{d}} \approx 2\times 10^{-4}~{\mathrm{s}}^{-1}$, and $N_{\mathrm{crit}} \approx 4\times 10^{17}$.}
\label{lasing}
\end{center}
\end{figure} 
\end{document}

\title{Supplementary information:\\ 
Coherently amplifying photon production from vacuum with a dense cloud of accelerating photodetectors}

\author{Hui Wang}

\author{Miles Blencowe}

\address{Department of Physics and Astronomy, Dartmouth College, Hanover, New Hampshire
03755, USA }

\maketitle

\section*{\label{a} \uppercase{Supplementary note 1: Derivation of the RWA Hamiltonian from the field action}}
In the following, we derive the RWA Hamiltonian (3) starting from the relativistic cavity field-coupled-$N$ detector action (1). 

The starting model action for the cavity field-$N$ detector system is given by
\begin{eqnarray} 
S & = & -\int d^{3+1}x \frac{1}{2} \partial_{\mu} \Phi\partial^{\mu}\Phi + \sum_{i=1}^N \int d\tau_i \left\{\frac{m_0}{2} \left[\left(\frac{d Q_i}{d\tau_i}\right)^2 - \omega_{{\mathrm{d}}0}^2Q_i^2 \right] -\frac{g}{4!} Q_i^4 \right\}\cr
&&+ \lambda_0 \sum_ {i=1}^N\int d\tau_i \int d^{3+1}x Q_i(\tau_i) \Phi(t,\vec{r}) \delta^{3+1} \left[x^{\mu}-z_i^{\mu}(\tau_i)\right],
\label{action}
\end{eqnarray}
where the $i$th detector's worldline is $z^\mu_i(t) = (t, L_x/2 +A \cos\left(\Omega_\mathrm{m} t+\phi_i\right), L_y/2, L_z/2)$, and $\tau_i$ denotes the $i$th detector's proper time.  With the relation $S =\int dtL$, the system Lagrangian in the laboratory frame is
\begin{eqnarray}
L&=&-\int d^3r \frac{1}{2} \partial_{\mu} \Phi\partial^{\mu} \Phi + \sum_{i=1}^N \left\{\frac{m_0}{2} \left[\frac{dt} {d\tau_i} \left(\frac{d Q_i}{dt}\right)^2 - \frac{d\tau_i} {dt} \omega_{{\mathrm{d}}0}^2 Q_i^2\right] - \frac{d\tau_i} {dt} \frac{g}{4!} Q_i^4 \right\} \cr
&& + \frac {\lambda_0}{c} \sum_ {i=1}^N \frac{d\tau_i}{dt} \int d^3r Q_i(t) \Phi(t,\vec{r}) \delta^3(\vec{r} - \vec{r}_i(t)),
\label{lagrangian}
\end{eqnarray}
where the Lorentz factor ${d\tau_i}/ {dt} = \sqrt{1-\xi^2 \sin^2 \left(\Omega_\mathrm{m} t+\phi_i\right)}$ with $\xi = \Omega_\mathrm{m} A /c$. Performing the Legendre transformation on the Lagrangian (\ref{lagrangian}),
we obtain the following cavity field-detector system Hamiltonian:
\begin{eqnarray} 
H & = & \int d^3r \frac{1}{2} \left[ \Pi^2 + (\vec{\triangledown} \Phi)^2 \right] + \sum_{i=1}^N \frac{d\tau_i} {dt} \left[\frac{P_i^2} {2m_0} + \frac{m_0} {2} \omega_{{\mathrm{d}}0}^2 Q_i^2 + \frac{g}{4!} Q_i^4 \right] \cr
&& - \frac {\lambda_0}{c} \sum_ {i=1}^N \frac{d\tau_i}{dt} \int d^3r Q_i(t) \Phi(t,\vec{r}) \delta^3(\vec{r} - \vec{r}_i(t)),
\label{hamiltonian1}
\end{eqnarray}
where $\Pi(t, \vec{r}) = \dot{\Phi} (t, \vec{r})/c^2$ and $P_i(t)=m_0 \dot{Q}_i(t) {dt}/{d\tau_i}$ are the momenta conjugate to $\Phi(t, \vec{r})$ and $Q_i(t)$, respectively. The first term in Eq.~(\ref{hamiltonian1}) is the free cavity scalar field Hamiltonian, the second term is the free detector Hamiltonian, and the third term is the interaction Hamiltonian.

Quantizing by replacing the position and momentum coordinates with their corresponding operators, we assume weak anharmonic potential energy terms for the detector degrees of freedom, so that the free detector energy eigenvalues $E_n,\, n=0,1,2,\dots$, can be approximated using first order perturbation theory. The approximate energy eigenstates for the free detectors are then harmonic oscillator Fock states: $|n_0, n_1,\dots,n_N\rangle$, $n_i=0, 1, 2,...$. With the parametric resonance condition applied between a given detector's centre of mass oscillation frequency, cavity mode frequency, and transition frequency between the ground and excited detector's energy levels (see below), we can truncate the Hilbert space of each detector to its two lowest energy eigenstates $|n_i\rangle,\, n_i=0, 1$, associated with the ground and first excited energy eigenvalues $E_0,\, E_1$, respectively. Therefore, defining $\tilde{\omega}_\mathrm{d} = (E_1-E_0)/\hbar$, the free detectors' position operators and free Hamiltonian are expressed in terms of Pauli operators:
\begin{eqnarray} 
Q_i &=& \sqrt{\frac{\hbar} {2m_0\omega_{{\mathrm{d}}0}}} \sigma_i^x, \\
H_\mathrm{det} &=& \frac{\hbar \tilde{\omega}_\mathrm{d}} {2} \sum_{i=1}^N \frac{d\tau_i} {dt} \sigma_i^z.
\label{detector}
\end{eqnarray}

We consider a 3D cavity box with side lengths $L_x$, $L_y$, and $L_z$, and impose the Dirichlet boundary conditions $\Phi(t, 0, y, z) = \Phi(t, L_x, y, z) = 0$, $\Phi(t, x, 0, z) = \Phi(t, x, L_y, z) = 0$, and $\Phi(t, x, y, 0) = \Phi(t, x, y, L_z) = 0$. The cavity quantum field operator can be decomposed in terms of classical normal mode solutions and associated  creation and annihilation operators $a_{\vec{n}}$, $a_{\vec{n}}^{\dag}$ as follows:
\begin{eqnarray}
\Phi(t, \vec{r}) &=& \sum_{\vec{n}} \sqrt{\frac{2\hbar c}{2 L_x L_y L_z |k_{\vec{n}}|}} \sin\left({k_n}_x x\right) \sin\left({k_n}_y y\right) \sin\left({k_n}_z z\right) \left( a_{\vec{n}}(t) + a_{\vec{n}}^{\dag}(t) \right),
\label{phi}
\end{eqnarray}
with ${k_n}_i = n_i \pi/ L_i$, $i = x, y, z$. Under the resonance condition (see below), the cavity field  is truncated to the single mode with  vector label $\vec{n}=(2,1,1)$. The corresponding mode frequency is $\omega_\mathrm{c}=c k_\mathrm{c}$, $k_\mathrm{c}=\sqrt{(2\pi/L_x)^2+(\pi/L_y)^2+(\pi/L_z)^2}$.

Substituting (S4-S6) into Hamiltonian~(\ref{hamiltonian1}) and dropping the subscript $2,1,1$ on $a_{2,1,1} ^{(\dag)}$ for notational convenience, we then obtain the single-mode, relativistic parametrically driven Dicke Hamiltonian (2):
\begin{eqnarray} 
H & = & \hbar \omega_\mathrm{c} a^{\dag} a + \frac{\hbar \tilde{\omega}_\mathrm{d}} {2} \sum_ {i=1}^N \frac{d\tau_i} {dt} \sigma_i^z + \hbar \tilde{\lambda} \sum_ {i=1}^N \frac{d\tau_i} {dt} \sin \left[k_\mathrm{c} A \cos(\Omega_\mathrm{m} t + \phi_i) \right] ({a} + {a}^{\dag}) \sigma_i^x,
\label{hamilton}
\end{eqnarray}
where $\tilde{\lambda} = \sqrt{2}\hbar \lambda_0 / \sqrt{m_0 \omega_{\mathrm{d}0} c L_x L_y L_z |\vec{k} _{2,1,1}|}$.

Hamiltonian (\ref{hamilton}) expressed in terms of  $J^z$, $J^{\pm}$, $\xi=\Omega_\mathrm{m} A/c$, and with the detectors' (TLSs') phases all set to zero ($\phi_i=0$), is as follows:
\begin{eqnarray} 
H & = & \hbar \omega_\mathrm{c} a^{\dag} a + \hbar  \tilde{\omega}_\mathrm{d} \frac{d\tau} {dt} J^z + \hbar \tilde{\lambda} \frac{d\tau} {dt} \sin \left[\frac {\omega_\mathrm{c} \xi} {\Omega_\mathrm{m}} \cos(\Omega_\mathrm{m} t) \right] (\hat{a} + \hat{a}^{\dag}) J^x,
\label{dicke}
\end{eqnarray}
where the reciprocal Lorentz factor is ${d\tau}/ {dt} = \sqrt{1-\xi^2 \sin^2 (\Omega_\mathrm{m} t)}$. Applying the Fourier series expansion to $d\tau/dt$ and the Jacobi-Anger expansion to the $\sin \left[{\omega_\mathrm{c}\xi} \cos(\Omega_\mathrm{m}t) /{\Omega_\mathrm{m}}\right]$ term, we obtain
\begin{eqnarray}
\frac{d\tau}{dt}=\sum_{n=0}^{\infty}(-1)^n\dbinom{\frac{1}{2}}{n}\dbinom{2n}{n}\left(\frac{\xi}{2}\right)^{2n}+2\sum_{n=1}^{\infty}\sum_{n^{\prime}=1}^n(-1)^{n-n^{\prime}}\dbinom{\frac{1}{2}}{n}\dbinom{2n}{n-n^{\prime}}\left(\frac{\xi}{2}\right)^{2n}\cos\left(2n^{\prime} \Omega_\mathrm{m}t\right),
\label{dtaudt}
\end{eqnarray}
\begin{eqnarray}
\sin\left[\frac{\omega_\mathrm{c}\xi}{\Omega_\mathrm{m}}\cos(\Omega_\mathrm{m}t)\right]= 2\sum_{n=0}^{\infty}(-1)^nJ_{2n+1}\left(\frac{\omega_\mathrm{c}\xi}{\Omega_\mathrm{m}}\right)\cos\left[(2n+1) \Omega_\mathrm{m}t \right],
\label{sintermeq}
\end{eqnarray}
where $J_{2n+1}(z)$ is a Bessel function of the first kind.
Keeping only terms up to second harmonics in $\Omega_\mathrm{m}$, Eqs. (\ref{dtaudt}) and (\ref{sintermeq}) become approximately
\begin{eqnarray}
&&\frac{d\tau}{dt}\approx D_0+D_2\cos(2\Omega_\mathrm{m}t),\label{dtapproxeq}\\
&&\frac{d\tau}{dt} \sin\left[\frac{\omega_\mathrm{c}\xi}{\Omega_\mathrm{m}} \cos(\Omega_\mathrm{m}t)\right] \approx C_1\cos(\Omega_\mathrm{m}t),
\label{sinapproxeq}
\end{eqnarray}
where the $\xi$ dependent $D_0$ and $D_2$ coefficients can be read off from Eq. (\ref{dtaudt}) and the  $\xi$, $\Omega_\mathrm{m}$ dependent coefficient $C_1 = 2J_1\left ({\omega_\mathrm{c}\xi}/ {\Omega_\mathrm{m}} \right)$ can be read off from Eq. (\ref{sintermeq}). Substituting Eqs. (\ref{dtapproxeq}), (\ref{sinapproxeq}) into Hamiltonian Eq.~(\ref{dicke}), we obtain
\begin{equation}
H=\hbar \omega_\mathrm{c}a^{\dag}a+\hbar\left[\omega_\mathrm{d}+ \tilde{\omega}_\mathrm{d} D_2\cos(2\Omega_\mathrm{m}t) \right] J^z + \hbar\tilde{\lambda} C_1\cos(\Omega_\mathrm{m} t)(a^{\dag}+a) (J^+ + J^-),
\label{seriesapproxhameq}
\end{equation}
where $\omega_\mathrm{d}= \tilde{\omega}_\mathrm{d} D_0$ is the renormalized detector oscillator frequency. Transforming to the rotating frame via the unitary operator $U_{\mathrm{RF}}(t)=\exp\left(i\omega_\mathrm{c}a^{\dag}at + i J^z \left[\omega_\mathrm{d}t + { \tilde{\omega}_\mathrm{d} D_2} \sin(2\Omega_\mathrm{m}t) /2\Omega_\mathrm{m} \right]\right)$, the cavity mode and detector annihilation operators pick up time-dependent phase terms as follows: 
\begin{eqnarray}
a(t)&\to& e^{-i\omega_\mathrm{c}t}a(t),\cr 
\sigma_i^-(t)&\to& e^{-i\left[\omega_\mathrm{d}t+\frac{ \tilde{\omega}_\mathrm{d} D_2} {2\Omega_\mathrm{m}}\sin(2\Omega_\mathrm{m} t)\right]}\sigma_i^-(t).
\end{eqnarray}
The system Hamiltonian (\ref{seriesapproxhameq}) then becomes  in the interaction picture:
\begin{equation}
H_I=\hbar \tilde{\lambda} C_1 \cos(\Omega_\mathrm{m}t) (e^{i\omega_\mathrm{c}t} a^{\dag} + e^{-i\omega_\mathrm{c}t}a) \left[e^{i\omega_\mathrm{d}t} e^{iB\sin(2\Omega_\mathrm{m}t)} J^+ + e^{-i\omega_\mathrm{d}t} e^{-iB \sin(2\Omega_\mathrm{m}t)} J^-\right],
\label{interaction}
\end{equation}
where $B= \tilde{\omega}_\mathrm{d} D_2/2\Omega_\mathrm{m}<1$.
Making use of the Jacobi-Anger expansion again such that
\begin{equation}
e^{\pm iB\sin(2\Omega_\mathrm{m}t)}\approx J_0\left(B\right)\pm 2i J_1\left(B\right)\sin(2\Omega_\mathrm{m}t),
\label{JAeq}
\end{equation}
and substituting Eq. (\ref{JAeq}) into Eq.~(\ref{interaction}),  we arrive at the following expression for the system Hamiltonian:
\begin{eqnarray}
H_I & \approx & \hbar \tilde{\lambda} C_1 \cos(\Omega_\mathrm{m}t) \left \{e^{i(\omega_\mathrm{c} + \omega_\mathrm{d})t} \left[J_0 \left(B\right) + 2i J_1 \left(B\right) \sin(2\Omega_\mathrm{m}t) \right] a^{\dag} J^+ \right. \cr
&&\left.+e^{-i(\omega_\mathrm{c}+\omega_\mathrm{d})t}\left[J_0\left(B\right)-2i J_1\left(B\right)\sin(2\Omega_\mathrm{m}t)\right]aJ^-\right. \cr
&&\left.+e^{i(\omega_\mathrm{c}-\omega_\mathrm{d})t}\left[J_0\left(B\right)-2i J_1\left(B\right)\sin(2\Omega_\mathrm{m}t)\right]a^{\dag}J^-\right. \cr
&&\left.+e^{-i(\omega_\mathrm{c}-\omega_\mathrm{d})t}\left[J_0\left(B\right)+2i J_1\left(B\right) \sin(2\Omega_\mathrm{m}t) \right]aJ^+ \right\}.
\label{interaction2}
\end{eqnarray}
\begin{figure}[ht]
\begin{center}
\includegraphics[width=0.8\textwidth]{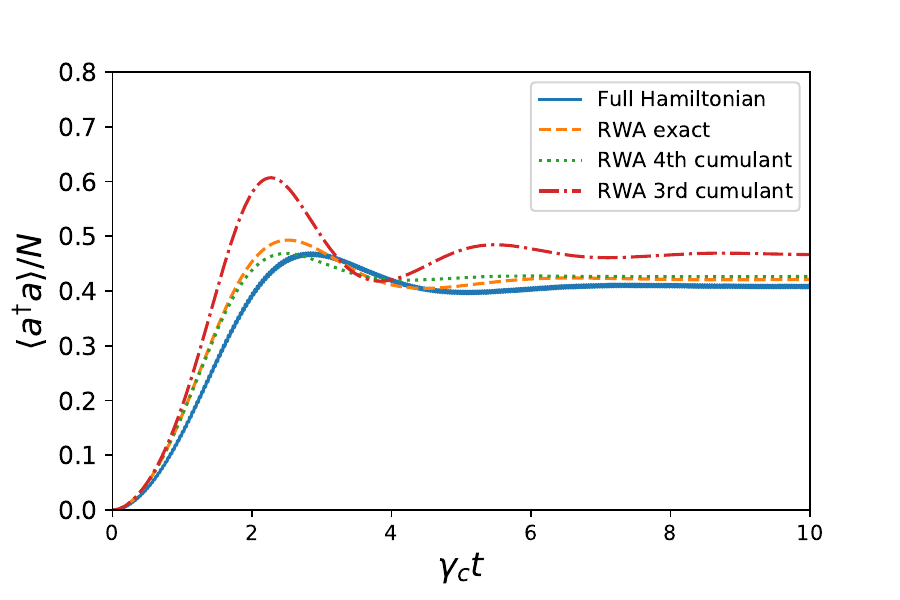} 
\caption*{{\bf Supplementary Figure 1. Dynamical evolution of the (scaled) cavity average photon number for seven TLSs  undergoing highly relativistic centre of mass oscillatory acceleration}. The TLSs are coupled to independent environments, and their centre of mass oscillation  phases $\phi_i$ are randomly chosen between 0 and $2\pi$. The solid line is for the relativistic, parametrically driven Dicke Hamiltonian (2) [Eq. (\ref{hamilton})], and the dashed line is for the RWA Hamiltonian (3) [Eq. (\ref{RWAHameq})], both calculated numerically using QuTiP \cite{johansson2013}; the dotted and dash-dotted lines are for the 4th and 3rd cumulant expansion approximations to the RWA Hamiltonian (3), respectively. The chosen example parameters are $\omega_\mathrm{c} = \omega_{\mathrm{d}0} = 1.0$, $\xi=0.8$, $\gamma_\mathrm{c}=\gamma_\mathrm{d}=0.02$, and $\lambda_0=0.0148$. The renormalized coupling $\lambda=0.01$, corresponding to $N_\mathrm{crit} = 1$.
\label{validity}
}
\end{center}
\end{figure}
Imposing the parametric resonance condition $\Omega_\mathrm{m}=\omega_\mathrm{c}+\omega_\mathrm{d}$ and combining the $\cos(\Omega_\mathrm{m}t)$ term with the first two terms within the braces, we retain the time-independent terms and drop the oscillating terms at integer multiples of $\Omega_\mathrm{m}$  (RWA).  As a result, we recover the approximate time independent Hamiltonian (3): 
\begin{equation}
H  \approx\hbar\lambda (a^{\dag} J^+ + aJ^-),
\label{RWAHameq}
\end{equation}
where we have dropped the subscript $I$ and the renormalized coupling constant $\lambda=\frac{1}{2} \tilde{\lambda} C_1 \left[J_0 \left ({\tilde{\omega}_\mathrm{d} D_2} /{2\Omega_\mathrm{m}}\right) -J_1 \left({\tilde{\omega}_\mathrm{d} D_2} /{2\Omega_\mathrm{m}}\right)\right]$. Supplementary Figure 1 compares the average photon number dynamics for the relativistic, parametrically driven Dicke Hamiltonian Hamiltonian (2) [Eq. (\ref{hamilton})] with the dynamics for the RWA Hamiltonian (3) [Eq. (\ref{RWAHameq})]. The plots are obtained by numerically solving for some example parameter values the Lindblad master equation (4) using the Quantum Toolbox in Python (QuTiP) \cite{johansson2013}.  
From the figure, we can see that the Lindblad master equation (4) with RWA Hamiltonian (3) [Eq. (\ref{RWAHameq})]  gives a good approximation to the average photon number dynamics for the  parametrically driven Dicke Hamiltonian (2) [Eq. (\ref{hamilton})], even for relativistic motion with $\xi\sim 1$. Furthermore, the dynamics is relatively insensitive to the individual TLS's phases $\phi_i$. 

Under nonrelativistic conditions $\xi\ll 1$ relevant for possible experimental realizations, we have $D_0\approx 1$, $D_2\approx 0$ (with $d\tau/dt\approx 1$), and $C_1\approx \omega_\mathrm{c}\xi/\Omega_\mathrm{m}$, so that the renormalized coupling is $\lambda\approx \tilde{\lambda} \omega_\mathrm{c}\xi/(2\Omega_\mathrm{m})$; the Lindblad master equation (4) with RWA Hamiltonian (3) [Eq. (\ref{RWAHameq})] gives an even more accurate approximation to the average photon number dynamics for the parametrically driven Dicke Hamiltonian Hamiltonian (2) [Eq. (\ref{hamilton})] in the nonrelativistic regime than for the relativistic regime.  

\section*{\label{b}\uppercase{Supplementary note 2: Cumulant expansion}}
In the following, we derive from the Lindblad master equation (4), approximate equations for the second moments of the cavity and TLSs degrees of freedom via a cumulant expansion.

We start from the dynamics of the cavity-TLS system as described by the Lindblad master equation:
\begin{eqnarray}
\dot{\rho} = -\frac{i}{\hbar} [H,\rho] + \gamma_\mathrm{c} \mathcal{L}_{a}[\rho] + \sum_{i=1}^N\gamma_\mathrm{d} \mathcal{L}_{\sigma_i^-} [\rho],
\label{meq}
\end{eqnarray}
where the Lindblad superoperators are defined by $\mathcal{L}_{A} [\rho] = A\rho A^{\dag} - \frac{1}{2} A^{\dag}A\rho - \frac{1}{2} \rho A^{\dag}A$. Consider the unitary transformation $G=\exp [i\theta (a^{\dag} a - J^z + N/2)]$, which leaves Hamiltonian (3) [Eq. (\ref{RWAHameq})] invariant. The master equation for the transformed system density matrix $\tilde{\rho} = G\rho G^{\dag}$  is given by
\begin{eqnarray}
\dot{\tilde{\rho}} &=& -\frac{i}{\hbar} G[H,\rho] G^{\dag}+ \gamma_\mathrm{c} G \mathcal{L}_{a}[\rho] G^{\dag} +  \gamma_\mathrm{d} \sum_{i=1}^N G\mathcal{L}_{\sigma_i^-} [\rho] G ^{\dag}\cr
&=& -\frac{i}{\hbar} [H,\tilde{\rho}] + \gamma_\mathrm{c} \mathcal{L}_{a}[\tilde{\rho}] + \gamma_\mathrm{d} \sum_{i=1}^N \mathcal{L}_{\sigma_i^-} [\tilde{\rho}] ,
\label{tildemeq}
\end{eqnarray}
where we have applied the transformation relations $G^{\dag} a G=e^{i\theta}a$ and $G^{\dag} \sigma_i^- G=e^{-i\theta} \sigma_i^-$. Assume that the system state is initialized as a product state of the individual ground states of the isolated cavity and $N$ isolated TLSs: $|\psi(0)\rangle = |0\rangle_c \otimes |-, -, \dots, -\rangle_s$. The fact that $\tilde{\rho}(0) = \rho(0)$ and the transformed equation (\ref{tildemeq}) coincides with Eq. (\ref{meq}), indicates that $\tilde{\rho}(t) = \rho(t)$ for the whole time range: $G$ is a symmetry of the cavity-TLS system and initial state. As a consequence, only operators that are invariant under the $G$ transformation give non-vanishing expectation values, and thus for the first and second moments, we need consider only the following non-zero moments: $\langle \sigma_1^z \rangle$, $\langle a^{\dag} a\rangle$, $\langle a\sigma_1^- \rangle$, $\langle a^{\dag} \sigma_1^+ \rangle$, and $\langle \sigma_1^+ \sigma_2^- \rangle$; all TLSs give the same moment values (a consequence of having the same assumed coupling and damping rates), which allows us to replace $\sigma_i^{z(+,-)}$ and $\sigma_i^+ \sigma_j^-$ ($i\neq j$) with $ \sigma_1^{z(+,-)}$ and $\sigma_1^+ \sigma_2^-$, respectively. We have verified numerically for a few $N$ TLSs using QuTiP \cite{johansson2013} that the following noninvariant moments that would otherwise appear in the moment equations vanish, i.e., $\langle a^2 \rangle = \langle aJ^+ \rangle = \langle J^+ J^+ \rangle=0$. 

With Eq.~(\ref{meq}), the dynamical differential equations for the  second moments can be obtained through a cumulant approximation \cite{Kubo}.  These equations also include the nonvanishing third moment terms $\langle a^{(\dag)} \sigma_2^{\pm} \sigma_1^z\rangle$ and $\langle a^{\dag} a \sigma_1^z\rangle$. The usual way to approximate these third moments is to set to zero the corresponding third order cumulants and obtain an approximate expression involving a sum of products of first and second order moments. However, we instead utilize an alternative cumulant approximation that is found to be more accurate. To proceed, we first rewrite $\sigma_1^z$ through the identity $\sigma_1^z = 2\sigma_1^+ \sigma_1^- -1$, and substitute into the third moments to obtain the sum of a fourth moment term and one second moment term (note that alternative substitutions such as $\sigma_1^z = 1-2\sigma_1^- \sigma_1^+$ give poorer approximations than the former, normal ordered choice).  By approximating the fourth moments through setting the fourth cumulants to zero, we obtain the following improved third-moment approximations:
\begin{eqnarray}
\langle a \sigma_2^- \sigma_1^z\rangle = \langle a \sigma_2^-\rangle \langle \sigma_1^z \rangle + 2\langle a \sigma_1^-\rangle \langle \sigma_2^- \sigma_1^+\rangle,\cr
\langle a^{\dag} \sigma_2^+ \sigma_1^z\rangle = \langle a^{\dag} \sigma_2^+\rangle \langle \sigma_1^z \rangle + 2\langle a^{\dag} \sigma_1^+\rangle \langle \sigma_1^- \sigma_2^+\rangle,\cr
\langle a^{\dag} a \sigma_1^z\rangle = \langle a^{\dag} a\rangle \langle \sigma_1^z \rangle + 2\langle a^{\dag} \sigma_1^+\rangle \langle a \sigma_1^-\rangle.
\label{fourthcumu}
\end{eqnarray}

The resulting approximate differential equations for the nonvanishing second moments and $\langle \sigma_1^z \rangle$ are
\begin{eqnarray}
\frac{d}{dt} \langle a^{\dag} a\rangle &=& -\gamma_\mathrm{c} \langle a^{\dag} a\rangle + 2iN\lambda \langle a\sigma_1^- \rangle, \cr
\frac{d}{dt} \langle a\sigma_1^- \rangle &=& -\frac{\gamma_\mathrm{c}+\gamma_\mathrm{d}} {2} \langle a\sigma_1^- \rangle - i\lambda \left[(N-1) \langle \sigma_1^+ \sigma_2^- \rangle-\langle a^{\dag} a\rangle \langle \sigma_1^z \rangle - 2\langle a \sigma_1^-\rangle \langle a^{\dag} \sigma_1^+\rangle + \frac{1-\langle \sigma_1^z \rangle}{2} \right],\cr
\frac{d}{dt} \langle a^{\dag} \sigma_1^+ \rangle &=& -\frac{\gamma_\mathrm{c}+\gamma_\mathrm{d}} {2} \langle a^{\dag} \sigma_1^+ \rangle + i\lambda \left[(N-1) \langle \sigma_1^+ \sigma_2^- \rangle-\langle a^{\dag} a\rangle \langle \sigma_1^z \rangle - 2\langle a \sigma_1^-\rangle \langle a^{\dag} \sigma_1^+\rangle + \frac{1-\langle \sigma_1^z \rangle}{2} \right],\cr
\frac{d}{dt} \langle \sigma_1^+ \sigma_2^- \rangle &=& -\gamma_\mathrm{d} \langle \sigma_1^+ \sigma_2^- \rangle - 2i\lambda \left[ \langle a \sigma_1^-\rangle \langle \sigma_1^z \rangle + 2\langle a\sigma_1^- \rangle \langle \sigma_1^+ \sigma_2^- \rangle \right],\cr
\frac{d}{dt} \langle \sigma_1^z \rangle &=& -\gamma_\mathrm{d} (\langle \sigma_1^z \rangle +1) + 4i\lambda \langle a\sigma_1^- \rangle.
\label{differential}
\end{eqnarray}
The above nonlinear dynamical equations must be solved numerically (see Supplementary Figure 1), although an approximate analytical solution for the average cavity photon number can be obtained in the long time limit steady state, defined by setting to zero the time derivatives of the moments. In particular, we obtain $\langle a^{\dag} a\rangle\approx N\gamma_\mathrm{d}/ (2\gamma_\mathrm{c})$ when $N\gg N_\mathrm{crit}$, where $N_\mathrm{crit}= \gamma_\mathrm{c} \gamma_\mathrm{d}/(4\lambda^2)$.

If we instead assume the usual, less accurate second moment approximation where the third order cumulants are set to zero, then the second terms on the right hand side of Eqs.~(\ref{fourthcumu}) are absent as well as the corresponding terms quadratic in the second moments in Eq. (\ref{differential}). This then enables the following general analytic solution to be obtained for the average cavity photon number in the long time limit:
\begin{eqnarray}
\langle a^{\dag}a\rangle = \frac{N\gamma_\mathrm{d} \{(N - N_\mathrm{crit}) (\gamma_\mathrm{c} + \gamma_\mathrm{d}) - 2 \gamma_\mathrm{c} + \sqrt{(N - N_\mathrm{crit})^2 (\gamma_\mathrm{c} + \gamma_\mathrm{d})^2 + 4 \gamma_\mathrm{c} \left[(N + N_\mathrm{crit}) (\gamma_\mathrm{c} + \gamma_\mathrm{d}) - \gamma_\mathrm{c} \right]} \}}{4 \gamma_\mathrm{c} \left[N (\gamma_\mathrm{c} + \gamma_\mathrm{d}) - \gamma_\mathrm{c} \right]}.
\label{anaindep}
\end{eqnarray}
In the limit $N,\, N_\mathrm{crit}\gg 1$, Eq.~(\ref{anaindep}) simplifies approximately to
\begin{eqnarray}
\langle a^{\dag}a\rangle \approx\begin{cases}
\frac{N \gamma_\mathrm{d}} {N_\mathrm{crit} (\gamma_\mathrm{c} + \gamma_\mathrm{d})} & N\ll N_\mathrm{crit} \\ \frac{N\gamma_\mathrm{d}} {2\gamma_\mathrm{c}} & N\gg N_\mathrm{crit}. \end{cases}
\label{cumlimit}
\end{eqnarray}
While the more accurate, vanishing fourth cumulant approximation does not give a  simple  general analytic expression like Eq. (\ref{anaindep}), it nevertheless yields the same approximate expressions as (\ref{cumlimit}) for $N,\,N_{\text{crit}}\gg 1$.  Note that the vanishing fourth cumulant approximation gives a more accurate approximation than the vanishing third cumulant approximation for the full dynamical evolution of the average cavity photon number, including the first burst peak dynamics  (see Supplementary Figure 1). 
%\section{Wigner function snapshots}
%\begin{figure}[ht]
%\begin{center}
%\includegraphics[width=0.8\textwidth]{wigner.pdf} 
%\caption{\label{wigner}Wigner function snapshots corresponding to the time instants indicated in Fig. 2b.}
%\end{center}
%\end{figure}

\vspace{0.5cm}
%\noindent{\large\textbf{Supplementary References}}\\